\documentstyle[twoside,fleqn,espcrc2,epsfig]{article}
\def\beq{\begin{equation}}
\def\eeq{\end{equation}}
\def\barr{\begin{eqnarray}}
\def\earr{\end{eqnarray}}
\def\lsim{\raise0.3ex\hbox{$\;<$\kern-0.75em\raise-1.1ex\hbox{$\sim\;$}}}
\def\gsim{\raise0.3ex\hbox{$\;>$\kern-0.75em\raise-1.1ex\hbox{$\sim\;$}}}
\def\lg{\raise0.3ex\hbox{$\;>$\kern-0.75em\raise-1.1ex\hbox{$<\;$}}}
\def\equiapp{\raise0.3ex\hbox{$\;\sim$\kern-0.75em\raise-1.1ex\hbox{$=\;$}}}

\newcommand{\AmS}{{\protect\the\textfont2
  A\kern-.1667em\lower.5ex\hbox{M}\kern-.125emS}}

\title{Dimensional Deconstruction and Neutrino Physics}

\author{K.R.S. Balaji 
\address {Department of Physics, McGill University, Montr\'eal, Queb\'ec, 
H3A 2T8, Canada}
\address{Email: balaji@hep.physics.mcgill.ca}}

\begin{document}
\begin{abstract}
We present a simple observation for neutrino mixings and masses which arises
naturally in dimensional deconstruction models. There are two essential ingredients
of such models: (i) the presence of a symmetry mediated by the link fields which
results in the  neutrino mixings to be maximal; and (ii) a deconstruction 
scale which for large values gives rise to a small neutrino mass, similar in feature 
to the seesaw mechanism.

\end{abstract}

\maketitle
\section{Introduction}

The current neutrino experiments have confirmed both the solar and atmospheric 
neutrino anomalies \cite{sksno}. There exist a large
class of neutrino models (see e.g. \cite{barr2000}) which explain the
anomaly and are yet not excluded. On the other hand, more general model 
dependent features require certain restrictions on the neutrino
spectra (see e.g. \cite{bala2000}). With in the framework of
dimensional deconstruction \cite{ark001,hill001} it is interesting 
to examine the Yukawa structure of an ultra-violet
complete theory in its infra-red limit.  It was first shown for neutrinos, that
by employing non-abelian symmetries on a latticized $S^1/Z_2$ orbifolding, 
dimensional deconstruction could lead to an agreement with data \cite{seidl003}. 
In a much simpler set-up \cite{bala}, 
we will show that the light neutrino mass (with maximal
mixing) is an outcome of deconstruction which projects out the seesaw operator
\cite{yana79}. 
   
\section{General frame work: the two-site model}
\label{twosite}
\begin{figure}
\begin{center}
\includegraphics*[bb = 228 646 355 717]{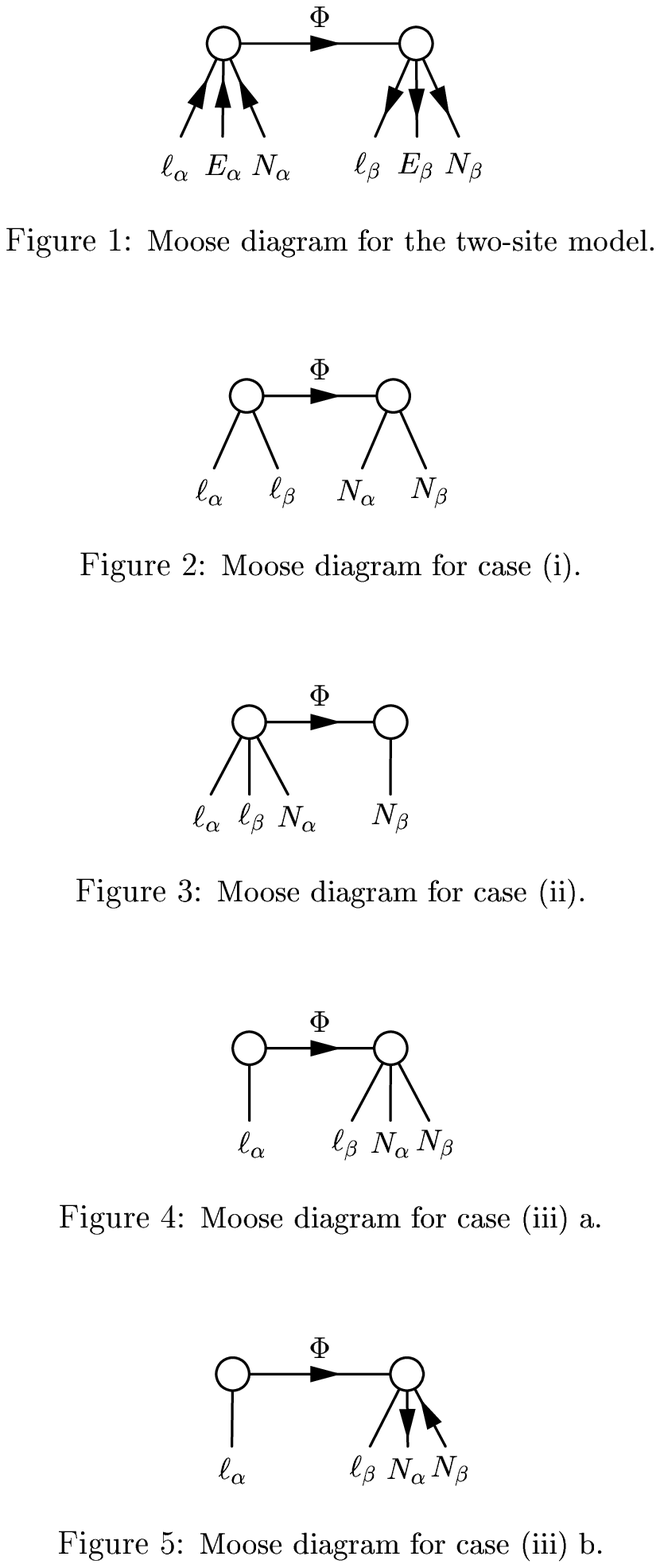}
\end{center}
\vspace*{-13.3mm}
\caption{\small{Moose diagram for the two-site model.}}
  \label{fig:twosites}
\end{figure}

We first present our basic two-site model which reproduces the seesaw 
operator along with maximal leptonic mixings in the light neutrino sector. 
Furthermore, we illustrate the twin combination 
({\it i.e.}, maximal mixing and small mass) is natural to this set up. In 
other words, we elucidate how our simple toy model makes a choice
with respect to the form of the Yukawa interactions.
Let us begin by considering a 
$G = G_{SM}\times SU(m)_1\times SU(m)_2$, where $G_{SM}$ denotes the 
usual standard model (SM) 
gauge group. 
\vspace*{-0.7mm}
The left-handed lepton doublets are denoted by 
$\ell_\alpha=(\nu_{\alpha L},\:e_{\alpha L})^T$ and the corresponding 
right-handed charged leptons by $E_\alpha$ and $e$. The leptons
$\ell_\alpha$ and $E_\alpha$ transform as $\overline{m}_1$ under $SU(m)_1$ and 
$\ell_\beta$ and $E_\beta$ transform as $m_2$ under $SU(m)_2$. To complete the
particle content, we add the right-handed neutrinos 
$N_{\alpha}$ and $N_{\beta}$ where, $N_{\alpha}$ transforms as 
$\overline{m}_1$ under $SU(m)_1$ while $N_{\beta}$ transforms as $m_2$ under 
$SU(m)_2$. The scalar link field $\Phi$ connects as the bi-fundamental
representation $(m_1,\overline{m}_2)$ the neighboring $SU(m)_i$ groups.
This field theory is summarized by a moose diagram shown in 
Fig.~\ref{fig:twosites}. We follow the conventions for the arrows as 
used in \cite{ark001}.
The corresponding renormalizable Yukawa interaction for the neutrino  
\beq\label{Yukawa1}
 L_Y=Y_{\alpha}\overline{\ell_\alpha}\tilde H N_{\alpha}
 +Y_{\beta}\overline{\ell_\beta}\tilde{H}N_{\beta}+
f\overline{N_{\alpha}^c}
 \Phi N_{\beta}
 +{\rm h.c.}~,
\eeq
where as usual, $\tilde{H}=i\sigma^2 H^\ast$ and 
$Y_{\alpha},Y_{\beta},f$ are complex Yukawa couplings.
 The kinetic term for the link field is 
\barr
L_{\Phi} &=& (D_\mu \Phi)^\dagger D^\mu \Phi~,\nonumber\\ 
D_\mu\Phi &=& (\partial_\mu -ig_1A_{1\mu}^a T_a+ig_2A_{2\mu}^aT_a)
\Phi~.
\label{phi}
\earr
The gauge fields are $A_{i\mu}^a$ $(i=1,2)$ with the generators ($T_a$) and 
dimensionless couplings, $g_1$ and $g_2$. It is important to note the 
absence of bare Dirac 
and Majorana mass terms which is due to invariance
under the group $G$. Once the link field develops a VEV, 
$\langle \Phi\rangle=M_x$ (also taken to be the scalar mass), the 
deconstruction scale breaks the $SU(m)_1\times SU(m)_2$ 
symmetry down to the diagonal $SU(m)$, thereby eating one adjoint
Nambu-Goldstone multiplet in the process. The deconstruction scale
corresponds to lattice spacing $a\sim 1/M_x$. The relevant Majorana mass 
matrix in the basis spanned by the neutrino fields
$(\nu_{\alpha L},\nu_{\beta L},N_{\alpha}^c, N_{\beta}^c)$ takes the form
\begin{equation}
\left( \begin{array}{cccc}
  0 & \!0\! & \!Y_{\alpha}\epsilon\! & 0  \\[-.5ex]
  0 & \!0\! & \!0\! & Y_{\beta}\epsilon \\[-.5ex]
  Y_{\alpha}\epsilon & \!0\! &\!0\! & fM_x\\ [-.5ex]
 0 & \!Y_{\beta}\epsilon\! &\!fM_x\! & 0\\
\end{array} \right), 
\label{matrix}
\end{equation}
where we have $\epsilon\equiv\langle H\rangle\simeq 10^2\:{\rm GeV}$ as the
electroweak scale.  At the scale $\epsilon$ the mass structure in 
(\ref{matrix}) results in a light Majorana neutrino mass 
\begin{equation}
M_\nu=\Big (\frac{Y_{\alpha} Y_{\beta}}{f}\Big) \frac{\epsilon^2}{M_x}~,
\label{matrix2}
\end{equation}
along with maximal mixing.
Thus, starting for shorter length scales, $r\ll a$, where we 
retain a completely renormalizable massless four-dimensional theory, 
for length scales 
$r \gg a \sim 1/M_x$, the mass matrix in (\ref{matrix2}) is identical to the
one which arises from the usual dimension-five operator of the type
$\sim \nu\nu HH$. In addition, the infra-red limit naturally leads to 
maximal mixings between the two active neutrino flavors. This feature of 
maximal mixings can be easily understood once we realize that in 
our setup, the link field, $\Phi$, mediates a symmetry between each of the 
fermions $(N_{\alpha,\beta})$ placed at different lattice sites; which is also
reflected in the resulting mass matrix 
for $N_{\alpha,\beta}$ as observed in (\ref{matrix}). 
This is retained after symmetry breaking 
(when the fermion masses are generated), as there exists the diagonal subgroup 
$SU(m)$ which respects the symmetry such that the $\Phi$ field would 
transform as $(m,\overline m)$. In the gauge sector, this unbroken 
symmetry corresponds to the presence of a zero mode. This can be seen in
the kinetic term in (\ref{phi}) which gives a mass squared matrix to 
the gauge bosons 
\begin{equation}
M^2_G \sim\langle\Phi\rangle^2 (g_1A^a_{1\mu} - g_2A^a_{2\mu})^2~.
\label{gbmm}
\end{equation}
In the basis $A^a_{1\mu}$ and $A^a_{2\mu}$, we have
\begin{equation}
M^2_G \sim \langle\Phi\rangle^2
\left( \begin{array}{cc}
  g_1^2 &-g_1g_2   \\[-.2ex]
  -g_1g_2& g_2^2 \\
 \end{array} \right), 
\label{mg}
\end{equation}
with the zero mode wave function 
\begin{equation}\label{eigenstate}
 A^{a(0)}_\mu=\frac{1}{\sqrt{g_1^2+g_2^2}}(g_2 A_{1\mu}^a+g_1A^a_{2\mu})~.
\end{equation}
The Dirac sector of the model remains diagonal due to the nature of this
construction while maximal mixings 
are introduced solely from the heavy Majorana sector. As a passing remark, 
the spectrum of the resulting
light neutrinos are in opposite $\mathcal{CP}$ parities and there is no explicit 
leptonic $\mathcal{CP}$ violation.

\section{An example phenomenology}
\label{threesite}
\begin{figure}
\begin{center}
\includegraphics*[bb = 235 529 349 639]{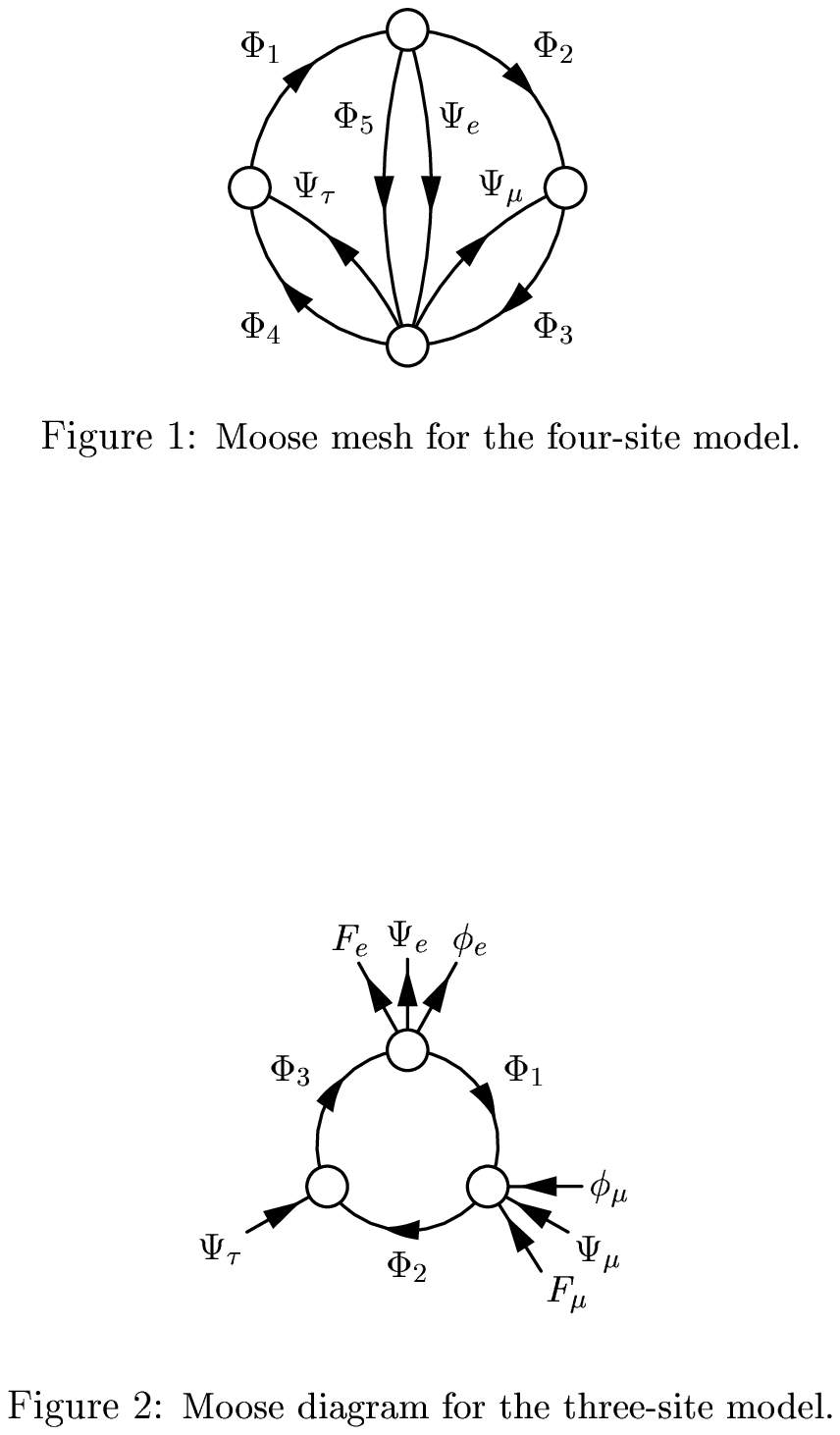}
\end{center}
\vspace*{-13.3mm}
\caption{\small{Moose mesh for the four-site model.}}
  \label{fig:moosemesh}
\end{figure}

We examine now a generalization to the case of a moose mesh.
Consider a $\Pi_{i=1}^4SU(m)_i$ gauge theory containing five scalar
link variables $\Phi_i$ $(i=1,\ldots,5)$ which transform under the $SU(m)$
gauge groups as
$\Phi_1\subset (m_1,\overline{m}_2)$,
$\Phi_2\subset (m_2,\overline{m}_3)$, $\Phi_3\subset (m_3,\overline{m}_4)$,
$\Phi_4\subset (m_4,\overline{m}_1)$, and $\Phi_5\subset (m_2,\overline{m}_4)$.
We will denote the flavor 
fields $\Psi_\alpha$ by the set
$\left\{\ell_\alpha,E_\alpha,N_\alpha\right\}$ of leptons. In our scheme, 
all the leptons associated with
$\Psi_\alpha$ carry the same $SU(m)_i$ charges and connect the different 
gauge groups as bi-fundamental representations. Hence, all 
fermions and scalars are treated identically as link variables;
see Fig.~\ref{fig:moosemesh}. 
Specifically, we make the choice $\Psi_e\subset (m_2,\overline{m}_4)$, 
$\Psi_\mu\subset (m_4,\overline{m}_3)$, and 
$\Psi_\tau\subset (m_4,\overline{m}_1)$. 
 After symmetry breaking and giving universal VEVs
$\langle\Phi_i\rangle\equiv M_x$ to the $\Phi_i$ fields, 
the structures of the Dirac and Majorana mass matrices take the form 
\begin{equation}\label{matrix1d}
M_D=\epsilon
\left(\begin{array}{ccc}
 Y_{e} &\lambda^2& \lambda^2\\
 \lambda^2&Y_{\mu}&\lambda^2\\
 \lambda^2&\lambda^2 &Y_{\tau}
\end{array}
\right)~,
\end{equation}
and 
\begin{equation}\label{matrix1m}
M_R=M_x
\left(\begin{array}{ccc}
 \lambda f_6 & f_1& f_2\\
 f_1&\lambda f_3&\lambda f_5\\
 f_2&\lambda f_5&\lambda f_4
\end{array}
 \right)~.
\end{equation}
\vspace*{-7mm}
In the above, $\lambda = M_x/\Lambda< 1$ and denotes the corrections to $M_D$ and
$M_R$ due to higher order terms (here up to dimension six) at scale $\Lambda$. 
Let us now consider an 
illustrative set of Yukawa values:
$f_1=-f_2=f_3=f_4=f_5\equiv f$ which leads to a familiar 
pattern of light neutrino mass matrix \cite{grim001}. For the choice 
$\lambda=0.22$, $Y_e=f$, $Y_\mu=f_6$, and the ratio 
$Y_{\mu}/Y_e = 2.5$, we 
obtain: (i) atmospheric mixing angle $\theta_{23}\simeq \pi/4$ (ii) 
$U_{e3}\simeq 0$. With the atmospheric splitting, 
$\Delta m_{\rm atm}^2=2.5\times 10^{-3}\:{\rm eV}^2$, we obtain a normal
neutrino mass hierarchy with the solar splitting
$\Delta m_\odot^2 \simeq 7.5 \times 10^{-5}\:{\rm eV}^2$ 
and the solar mixing, $\theta_{12}\simeq 32^\circ$ which corresponds to the
MSW LMA-I solution \cite{fogli002}.
The system predicts an effective neutrinoless double beta decay mass,
$m_{ee} \simeq 10^{-3}\:{\rm eV}$. 

 In summary, we demonstrate the potential richness of dimensional deconstruction
of latticized spatial dimensions for the flavor block of the SM.  
In this analysis, we have limited 
ourselves to describing the physics of a periodic lattice; where, we
take the two-site model as a periodic interval corresponding to a 
Brillouin zone. In the 
limit of a large lattice site model (of size $N\gg 1$), one can draw comparisons 
to a true extra-dimensional scenario (of radius $R$) along with the 
identifications 
to the five-dimensional gauges couplings, $g_5(y_i) \to \sqrt{R/N} g_i$. Here,
$y_i$ is taken to be the fifth coordinate. 

\section*{Acknowledgments}
 KB thanks the organizers of TAUP03 for a wonderful meeting and for the
local support. He also thanks a collaboration with G. Seidl and M. Lindner upon 
which this talk is based on. At McGill, KB is supported by funds from 
NSERC (Canada) and by Fonds de recherche sur la nature 
et les technologies of Qu\'ebec.

\end{document}